\documentstyle[12pt,aaspp4,epsf]{article}

\makeatletter
\def\U{4U1626--67}   \def\B{{\em BeppoSAX\/}}
\def\chidof{\mbox{$\chi^2_\nu$}}
\def\pmt#1#2{_{-#1}^{+#2}}
\def\@citex[#1]#2{\if@filesw\immediate\write\@auxout{\string\citation{#2}}\fi
  \def\@citea{}\@cite{\@for\@citeb:=#2\do
    {\@citea\def\@citea{;\penalty\@m\ }\@ifundefined
       {b@\@citeb}{{\bf ?}\@warning
       {Citation `\@citeb' on page \thepage \space undefined}}%
{\csname b@\@citeb\endcsname}}}{#1}}
\def\@cite#1#2{(#1\if@tempswa ; #2\fi)}
\def\newblock{\hskip .11em plus .33em minus .07em}
\makeatother

\lefthead{M. Orlandini et~al.}
\righthead{\B\ observations of \U}

\begin{document}

\title{\B\ observation of \U: Discovery of an absorption cyclotron resonance
feature}

\author{M. Orlandini, D. Dal~Fiume, F. Frontera\altaffilmark{1}}
\affil{Istituto Tecnologie e Studio Radiazioni Extraterrestri (TeSRE), C.N.R.
\\
via Gobetti 101, 40129 Bologna, Italy; orlandini@tesre.bo.cnr.it}
\and
\author{S. Del~Sordo, S. Piraino, A. Santangelo, A. Segreto}
\affil{Istituto Fisica Cosmica e Applicazioni all'Informatica (IFCAI), C.N.R.
\\
via La Malfa 153, 90146 Palermo, Italy}
\and
\author{T. Oosterbroek, A.N. Parmar}
\affil{Astrophysics Division, Space Science Department of ESA, ESTEC \\
Keplerlaan 1, 2200 AG Noordwijk, The Netherlands}

\altaffiltext{1}{Also Physics Dept., Ferrara University, via Paradiso 12, 44100
Ferrara, Italy}

\begin{abstract}
We report on an observation of the low-mass X--ray binary \U\ performed during
the \B\ Science Verification Phase. An absorption feature at $\sim 37$~keV,
attributable to electron cyclotron resonance, has been discovered in its pulse
averaged spectrum. The inferred neutron star magnetic field strength is
$3.2\cdot (1+z) \times 10^{12}$ G, where $z$ is the gravitational redshift. The
feature is deep and narrow and is resolved in both the broad-band fit and
in the ratio of observed counts to those seen from the Crab. The cyclotron
resonance energy is in good agreement with the empirical relation between
cyclotron energy and high energy cutoff, while its width is in agreement with
the expected Doppler broadening of thermal electrons at the cyclotron resonance
frequency. The broad-band 0.1--200 keV spectrum is well fit by a two-component
model: a $0.27\pm 0.02$ keV blackbody and a power law with a photon index of
$0.89\pm 0.02$. This is the first broad-band observation made after the change
from spin-up to spin-down that occurred in mid 1990: it confirms the
harder spectrum with respect to those observed in the 2--10 keV range.
\end{abstract}

\keywords{binaries: close --- stars: neutron --- pulsars: individual: \U\ ---
          X--rays: stars} 

\section{Introduction}

Following the first detection in Her X--1 \cite{576}, line features in the
10--110 keV energy range were detected in some X--ray pulsars \cite{407,375}.
These features are interpreted as being due to electron cyclotron transitions
in the $\ga 10^{12}$~G magnetic field of the neutron star \cite{407}. Because
the energy of the fundamental cyclotron harmonic, $E_{\rm cyc}$, is related to
the neutron star magnetic field strength in units of $10^{12}$~G, $B_{12}$, by
the relation $E_{\rm cyc} = 11.6\, B_{12}\cdot (1+z)^{-1}$~keV, where $z$ is
the gravitational redshift, the observation of cyclotron resonance features
(CRFs) in pulsar spectra gives a direct measurement of the magnetic field of
the neutron star. From a theoretical point of view CRFs are expected to be
visible as absorption features. This was shown {\em e.g.\/} by Nagel (1981)
\nocite{306} for any reasonable optical depth. The feature is not strictly due
to absorption, but rather to scattering of photons in the wings of the line. Up
to now CRFs have been observed mainly in the spectra of young high-mass binary
systems, with the notable exception of Her X--1 \cite{576}.

The 7.7~s X--ray pulsar \U\ is one of the few pulsators among LMXRBs. Its
timing history is characterized by a sudden change in its spin state that
occurred on 1990 June \cite{1548}. Before this date the source was steadily
spinning-up, while from the transition to the present time \U\ is
spinning-down. Besides coherent pulsation, \U\ also shows quasi-periodic
oscillations (QPOs) both in X--rays \cite{29} and in the optical \cite{1618},
with a centroid frequency of $\sim 40$ mHz.

The pulse-phase averaged spectrum of \U\ has been described in terms of a two
component model: an absorbed blackbody with a temperature $kT\sim 0.6$ keV and
a power law \cite{658}. The spectrum also exhibits a high energy cutoff at
$\sim 20$ keV, interpreted as due to a possible CRF at that energy \cite{617},
and an emission Ne line complex at $\sim 1$~keV \cite{393}. While the spin
transition did not affect the high energy part of the spectrum, within the
accuracy of previous measurements, it did affect the 2--10 keV part
\cite{1590}: the power law photon index changed from $\sim 1.6$ \cite{617} to
$\la 0.7$ \cite{393}, with a correspondingly significantly lower 2--10 keV
flux, that shows an overall fading by a factor four from the HEAO1/A2
measurement \cite{1548}.

\section{Observations}

The \B\ satellite is a program of the Italian Space Agency (ASI), with
participation of the Netherlands Agency for Aerospace Programs (NIVR), devoted
to X--ray astronomical observations in the broad 0.1--300 keV energy band.
The payload includes four Narrow Field Instruments (NFIs) and two
Wide Field Cameras. The NFIs consist of Concentrators Spectrometers
(C/S) with 3 units (MECS) operating in the 1--10 keV energy band \cite{1532}
and 1 unit (LECS) operating in 0.1--10 keV \cite{1531}, a High Pressure Gas
Scintillation Proportional Counter (HPGSPC) operating in the 3--120 keV energy
band \cite{1533} and a Phoswich Detection System (PDS) with four scintillation
detection units operating in the 15--300 keV energy band \cite{1386}.

During the Science Verification Phase a series of well known X--ray sources
were observed in order to check the capabilities and performances of the
instruments on-board \B. \U\ is one of these sources and it was observed by \B\
on 1996 August 6, from 14:41 to 21:50 UTC, and from 1996 August 9 00:10 to
August 11 00:00. Data were telemetred in direct modes, which provide complete
information on time, energy and, if available, position for each photon. The
net exposure times for the four NFIs were 31, 97, 43 and 52 ks for LECS, each
of the MECS units, HPGSPC and PDS, respectively. These differences are due to
the LECS only being operated in satellite night-time, rocking of collimators
(for HPGSPC and PDS) and different filtering criteria during passages in the
South Atlantic Geomagnetic Anomaly (SAGA) and before and after Earth
occultations. Furthermore, HPGSPC data from the August 6 observation were not
included because of telemetry problems. The spurious spikes ($\Delta t \la
2$~s) that affect the PDS data were also filtered. These spikes occur at a
frequency of few per orbit, can reach several hundreds counts per second and
are likely due to the local environmental particle background. Their spectrum
is generally soft, showing a strong cutoff at $\sim 40$ keV \cite{1620}. Their
presence can severely affect both a spectral and timing analysis, expecially
for moderately faint sources in the PDS bandpass, such as \U.

The data analysis of all the NFIs but the LECS was carried out in the XAS V2.0
framework \cite{1619}. For the LECS data the SAXLEDAS V1.4.0 was used
\cite{1650}. We used the 1997 September revisions of the response matrices for
all NFIs but LECS, for which we used the matrix generated by LEMAT V3.5.3.

\section{Spectral Analysis}

The analysis of the LECS data has been already presented \cite{1612}, and will
not be discussed here: we will include their result on the Ne line complex. For
the imaging instruments, the standard extraction radius of $8'$ was used to
derive the source events. The rocking collimated instruments (HPGSPC and PDS)
net counts were obtained with the standard background subtraction procedure,
described in details in Manzo et~al.\ (1997) and Frontera et~al.\ (1997). The
systematic error in PDS background subtraction is consistent with zero for
observations longer than 100 ks and negligible for sources whose flux is higher
than 1 mCrab, such as \U\ \cite{1620}.

\subsection{The X--ray continuum}

The  spectra of each \B\ NFIs were first fit separately in order to derive the
best model able to describe the broad-band spectrum. We find that a model
including low energy absorption, a blackbody, a power law and a high energy
cutoff adequately describes all the NFIs spectra but the PDS, yielding \chidof
s of 1.13 (81 degrees of freedom, dof) for LECS, 1.11 (540) for MECS, 1.29
(144) for HPGSPC, and 1.89 (56) for PDS. Most of the contribution to the high
\chidof\ for PDS is due to deviations at $\sim 35$ keV. We used this model to
describe the broad-band 0.1--200~keV spectrum. We included in the fit a
variable normalization for each of the NFIs (we treated separately each MECS
unit), in order to take into account for the known calibration uncertainties
between the instruments. As high energy cutoff we used the standard X--ray
pulsar model $\exp[(E_c - E)/E_f]$, where $E_c$ and $E_f$ are the cutoff and
folding energy, respectively \cite{303}. The smoother Fermi-Dirac cutoff
\cite{1584} did not adequately describe the high energy tail. After the
inclusion of the Ne line complex at 1 keV we obtained a best fit with a reduced
\chidof\ of 1.369 for 706 dof. The fit results are summarized in
Table~\ref{tab:fit}. The normalizations among NFIs (MECS2 as reference) are
1.04 for LECS, 1.05 for HPGSPC and 0.72 for PDS.

Our data do not show the presence of an Iron K-shell line in 6.4--6.9 keV: the
3$\sigma$ upper limit on its equivalent width is 21 eV, slightly more stringent
than the 33 eV value obtained by ASCA \cite{393}. The total 0.1--200~keV X--ray
luminosity is $7.7\times 10^{34}$ erg~s$^{-1}$~d$^2_{\rm kpc}$. The fluxes in
the bands 2--10, 10--60, and 10--200 keV are 1.7, 4.4, and  $4.5\times
10^{-10}$ ergs~cm$^{-2}$~s$^{-1}$, respectively.

\subsection{Cyclotron resonance}

From the analysis of the residuals we were led to add an extra component to our
spectral model. Both a Lorenzian \cite{1547} or a Gaussian in absorption
\cite{1172} at $\sim 35$ keV, interpreted as a CRF, improved the fit, yielding
\chidof s of 1.146 and 1.232 for 703 dof, respectively. An F-test shows that
the improvement is significant at 99.99\%. Adding this component improved
the fit to the 15--200 keV PDS spectrum, yielding a \chidof\ of 1.256 (53 dof;
probability of chance improvement $2.4\cdot 10^{-5}$). The Gaussian line FWHM
of $7\pm 2$ keV is consistent with the PDS energy resolution at the CRF energy,
$\sim 17.5$\%  \cite{1386}. Figure~\ref{fig:spectra} shows the best-fitting
spectral model with the residual CRF Gaussian line profile obtained by setting
the line normalization to zero. The residuals for each NFIs are also shown.

The two CRF models are equivalent: the probability of chance improvement from
the Lorenzian to the Gaussian model is 34\%. The different E$_{\rm CRF}$ found
with the two models is an effect due to the modeling: the Lorenzian model
approximates the fall-off of the spectrum by increasing the cutoff energy, and
shifting the cyclotron energy to values lower than those obtained by an
absorption Gaussian (compare, in Table~\ref{tab:fit}, the best fit continuum
parameters with the CRF line included with those obtained from the continuum
without the line). This is a known effect, already observed for the CRF in Her
X--1 \cite{1583} and Vela X--1 \cite{1581}.

Following the suggestion by Pravdo et~al.\ (1978) \nocite{617} of a possible
CRF at $\sim 19$ keV, we tried to include in our broad-band model a fundamental
CRF at this energy (and therefore to consider the $\sim 35$ keV CRF as the
second harmonic). This attempt was unsuccessful, both with and without the
high energy cutoff component.

In order to better characterize the \U\ CRF we examined the ratio between the
PDS and Crab count rate spectra. This ratio has the advantage of minimizing the
effects due to the detector response, and the uncertainties in the
calibrations. As we can see from the upper panel in Fig.~\ref{fig:crab-ratio},
there is an evident change of slope at $\sim 35$ keV. The same ratios performed
on the Her X--1 \cite{1583} and Vela X--1 \cite{1581} spectra show a change of
slope at $\sim 40$ and $\sim 57$ keV, their well known cyclotron energies.

To enhance the effect due to the CRF, we multiplied the \U/Crab Ratio by
the functional form of the Crab spectrum, {\em i.e.\/} a simple power law with
index equal to 2.1. In this way we emphasize deviations of the \U\ spectrum
from its continuum without making any assumption on its form. The result is
shown in the second panel of Fig.~\ref{fig:crab-ratio}, where it is evident
that deviations occur only at E $\ga 30$ keV.

Finally, in the lower panel we show the ratio between the previous function and
the \U\ continuum model, with $\alpha$, $E_c$ and $E_f$ taken from
Table~\ref{tab:fit}. In this way we enlarge all the effects due to line
features, although we introduce a model dependence. The absorption feature at
$\sim 38$ keV is evident. This feature can be fit between 25 and 50 keV with
both a Gaussian (shown in the figure) and a Lorenzian. The results are shown in
Table~\ref{tab:crf_fit} and are in agreement with those obtained from the
broad-band spectral fit shown in Table~\ref{tab:fit}. The difference in the
$\sigma_{\rm CRF}$ values is the effect due to the intrinsic energy resolution
of the PDS instrument. The inferred magnetic field strength at the neutron star
surface is therefore $(3.2\pm 0.1)\times 10^{12}\cdot (1+z)$~G, where $z$ is
the gravitational redshift.

\section{Discussion}

This is the first \U\ broad-band spectrum obtained after the  spin transition.
From our flux measurement we can see that the overall {\em bolometric\/} flux
of the source decreased by a factor 4.1 since the HEAO1/A2 measurements. We
find a power law index slightly higher than those found in narrow-band spectra
\cite{393,1612}. This is an effect due to the broad--band fit, in which we
simultaneously take into account both the low--energy power law and the
high--energy cutoff.

Our \U\ CRF measurement, together with its cutoff energy, fits the empirical
relation found in X--ray binary pulsars between these parameters: higher the
cyclotron resonance energy, higher the cutoff --- and therefore harder the
spectrum \cite{61,1286}. Our CRF measurement also is in agreement with the
expected electron Doppler broadening. Indeed, at the cyclotron resonance
frequency $\omega_c$, electrons at rest absorb photons of energy
$\hbar\omega_c$. For moving, thermal electrons the Doppler broadening
$\Delta\omega_D$ is predicted to be \cite{1614}
\begin{equation}
\Delta\omega_D = \omega_c \left( \frac{2kT_e}{m_ec^2} \right)^{1/2}
  |\cos\theta| 
\label{eq:deltaD}
\end{equation}
where $kT_e$ is the electron temperature, and $m_ec^2$ is the electron rest
mass. The angle $\theta$ measures the direction of the magnetic field with
respect to the line of sight. Outside the range $\omega_c\pm\Delta\omega_D$ the
cyclotron absorption coefficient decays exponentially, and other radiative
processes become important. From Eq.~\ref{eq:deltaD} and the CRF parameters
given in Table~\ref{tab:fit} we obtain a lower limit to the electron
temperature of $\sim 5$ keV, in reasonable agreement with the calculations of
self-emitting atmospheres of Harding et~al.\ (1984) \nocite{304}, according to
which the temperature for a column or slab with optical depth $\sim 50$
g~cm$^{-2}$ is 4--8$\times 10^7~^\circ$K, corresponding to $kT_e\sim 3.5$--7
keV.

Finally, we want to mention that if we assume the QPO frequency as due to the
beating between the pulse frequency and the Keplerian motion at the
magnetospheric radius, then the \U\ magnetic field strength is related to its
luminosity by the relation $B_{12}\sim 5.5\sqrt{L_{37}}$ \cite{29}, where
$L_{37}$ is the X--ray luminosity in units of $10^{37}$~erg~s$^{-1}$. Assuming
a source distance $5\la \rm d_{\rm kpc} \la 13$ \cite{1618}, this corresponds
to $2.4\la B_{12} \la 6.3$, in  good agreement with our measurement.

\begin{acknowledgements}
The authors wish to thank the \B\ Scientific Data Center staff, the
Nuova-Telespazio OCC personnel, the \B\ Mission Scientist L.\ Piro and the \B\
Mission Director R.C.\ Butler. This research has been funded in part by the
Italian Space Agency. We thank the anonymous referee for useful suggestions.
\end{acknowledgements}

\newpage

\begin{deluxetable}{lllll}
\tablewidth{0pt}
\tablecaption{Best-fit spectral parameters\ \tablenotemark{a} \label{tab:fit}}
\tablehead{
\multicolumn{2}{l}{Parameter} & \multicolumn{3}{c}{Value} \\
 \cline{3-5}
 & & \colhead{No Line} & \colhead{Gaussian} & \colhead{Lorenzian}}
\startdata
N$_{\rm H}$     & ($10^{21}$ cm$^{-2}$)      & $1.1\pm 0.2$           & $1.2\pm 0.2$           & $1.1\pm 0.2$\nl
$kT$            & (keV)                      & $0.28\pm 0.01$         & $0.29\pm 0.01$         & $0.30\pm 0.01$ \nl
$r_{\rm bb}$    & ($\times \rm d_{\rm kpc}$ km) & $1.5\pmt{0.2}{0.3}$ & $1.5\pm 0.2$           & $1.4\pm 0.2$ \nl
$\alpha$        &                            & $0.87\pm 0.01$         & $0.86\pm 0.01$         & $0.83\pm 0.01$ \nl
E$_c$           & (keV)                      & $21.1\pm 0.4$          & $19.6\pm 0.5$          & $29\pm 2$ \nl
E$_f$           & (keV)                      & $7.8\pm 0.3$           & $10.6\pm 0.7$          & $9\pmt{1}{2}$ \nl
\tablenotemark{b}\ \ I$_{\rm pow}$ &         & $1.06\pm 0.02$         & $1.05\pm 0.02$         & $1.01\pm 0.02$ \nl
E$_{\rm Ne}$    & (keV)                      & 1.05 \tablenotemark{c} & 1.05 \tablenotemark{c} & 1.05 \tablenotemark{c} \nl
$\sigma_{\rm Ne}$ & (keV)                    & 0.04 \tablenotemark{c} & 0.04 \tablenotemark{c} & 0.04 \tablenotemark{a} \nl
EW$_{\rm Ne}$   & (keV)                      & 48   \tablenotemark{c} & 48   \tablenotemark{c} & 48 \tablenotemark{c} \nl
E$_{\rm CRF}$   & (keV)                      & \nodata                & $37\pm 1$              & $33\pm 1$ \nl
$\sigma_{\rm CRF}$ & (keV)                   & \nodata                & $3\pm 1$               & $12\pmt{2}{3}$ \nl
EW$_{\rm CRF}$  & (keV)                      & \nodata                & $15\pm 3$              & \nodata \nl
$\tau_{\rm CRF}$&                            & \nodata                & \nodata                & $1.6\pm 0.3$ \nl
\chidof         &                            & 1.369 (706)            & 1.232 (703)            & 1.146 (703) \nl
\tablenotetext{a}{ Uncertainties at 90\% confidence level for a single parameter.}
\tablenotetext{b}{ Flux is in units of $10^{-2}$~photons~cm$^{-2}$~s$^{-1}$ at 1 keV.}
\tablenotetext{c}{ The Ne line complex parameters are taken from Owens et~al.\ (1997).}
\enddata
\end{deluxetable}

\newpage

\begin{deluxetable}{lll}
\tablewidth{0pt}
\tablecaption{Fit results on the \U\ CRF\ \tablenotemark{a} \label{tab:crf_fit}}
\tablehead{
\colhead{CRF Parameter} & \colhead{Gaussian}  & \colhead{Lorenzian}}
\startdata
E$_{\rm CRF}$ (keV)      & $38.6\pm 0.9$ & $38.1\pm 0.9$ \nl
$\sigma_{\rm CRF}$ (keV) & $ 5.0\pm 0.7$ & $ 8\pm 1$     \nl
\chidof\ (dof) & 1.048 (13)    & 2.725 (13)    \nl
\tablenotetext{a}{ Uncertainties are given at 90\% confidence level.}
\enddata
\end{deluxetable}

\newpage

\figcaption[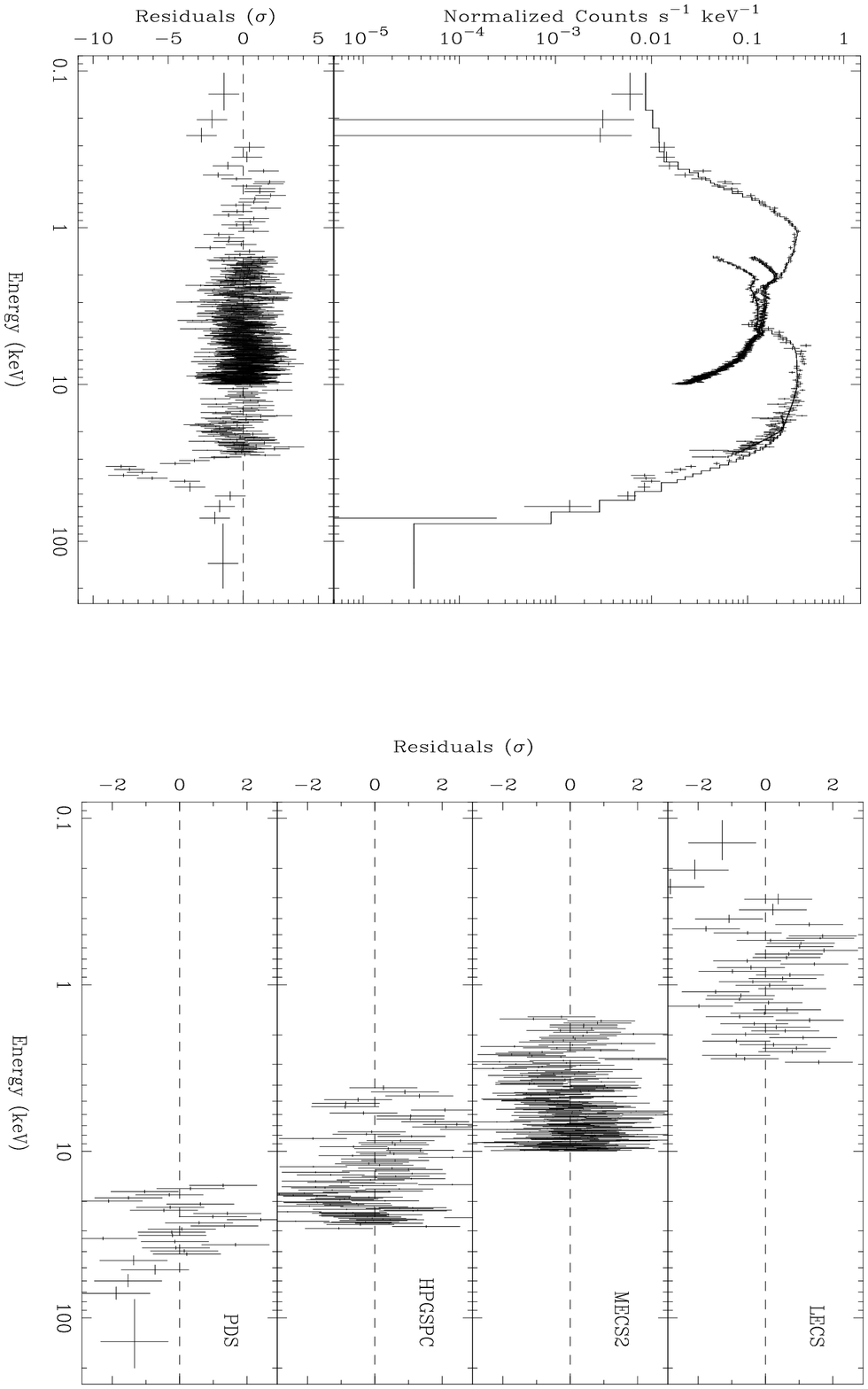]{The broad-band 0.1--200 keV spectrum of \U\ observed
by the \B\ NFIs. Left panel: count rate spectrum (crosses) and best-fitting
continuum model (histogram) with the Ne complex and absorption Gaussian
components (parameters from Table~1). 
The Gaussian normalization has been set to zero and the residuals from the
model are shown. Right panel: detailed NFIs residuals from the  broad-band
best-fit model (for clarity only one MECS unit is shown).
\label{fig:spectra}}

\figcaption[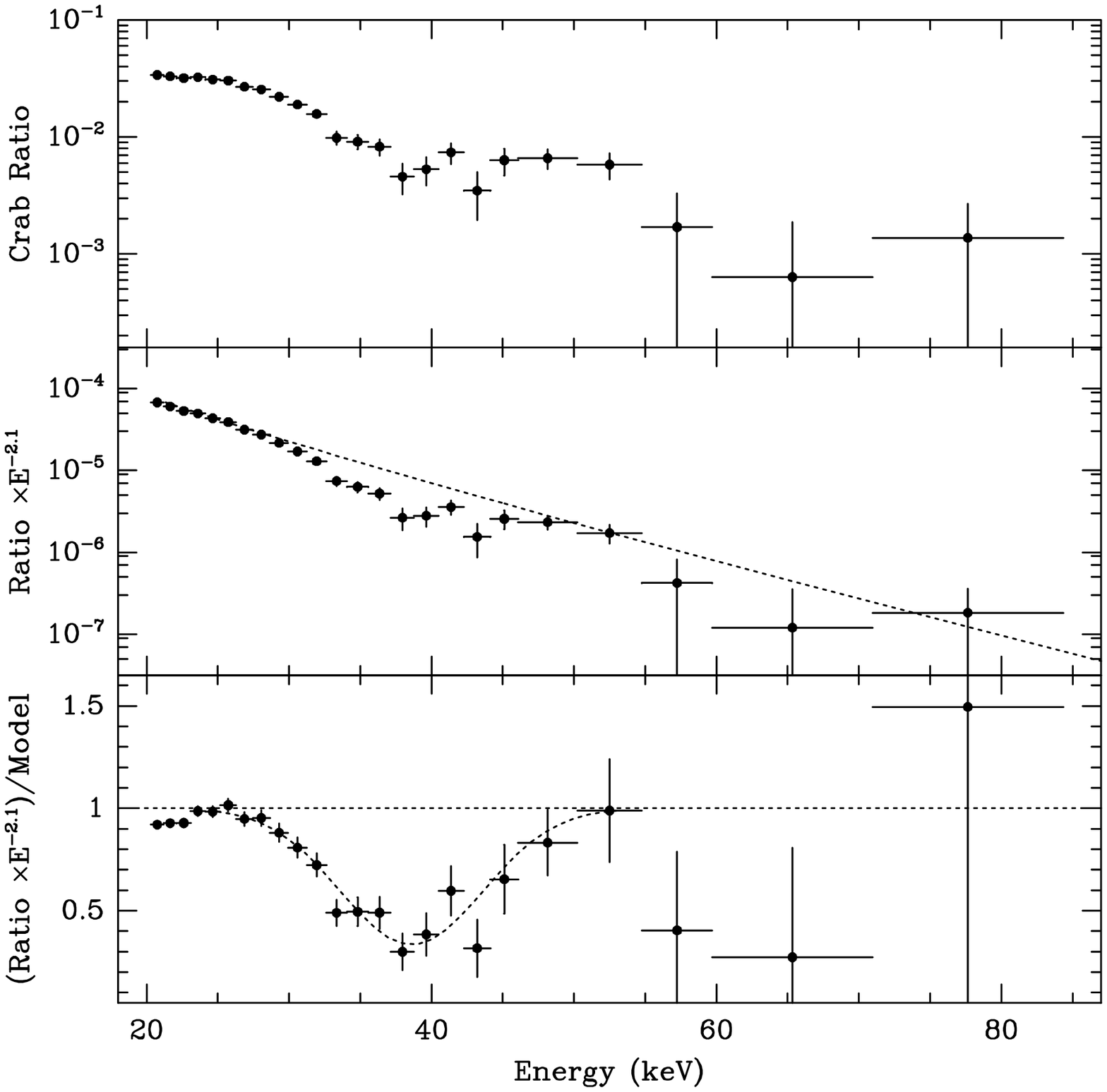]{Upper panel: Ratio between the PDS \U\ and Crab count
rate spectra. Note the change of slope at $\sim 35$ keV. Middle panel: the
\U/Crab ratio multiplied by E$^{-2.1}$, the functional form of the Crab
spectrum. Deviations from the continuum (dotted line) occur for E $\ga 30$ keV.
Lower panel: ratio between the (\U/Crab ratio $\times$ E$^{-2.1}$) and the
model of the \U\ continuum as derived from the broad-band fit. A Gaussian fit
to the CRF (dotted line) is also shown. \label{fig:crab-ratio}}

\newpage

\begin{figure}
\plotone{figure01.ps}
\end{figure}

\begin{figure}
\plotone{figure02.ps}
\end{figure}


\begin{thebibliography}{}

\bibitem[Angelini et~al.\ 1995]{393}
Angelini, L., et~al. 1995,
\newblock ApJ, 449, L41

\bibitem[Boella et~al.\ 1997]{1532}
Boella, G., et~al. 1997,
\newblock A\&AS, 122, 327

\bibitem[Chakrabarty 1998]{1618}
Chakrabarty, D. 1998,
\newblock ApJ, 492, 342

\bibitem[Chakrabarty et~al.\ 1997]{1548}
Chakrabarty, D., et~al. 1997,
\newblock ApJ, 474, 414

\bibitem[Chiappetti \& Dal~Fiume 1997]{1619}
Chiappetti, L., \& Dal~Fiume, D. 1997,
\newblock in {Proc. 5$^{\rm th}$ International Workshop on Data
  Analysis in Astronomy}, ed. L. Scarsi \& M.C. Maccarone (Singapore: World
  Scientific)

\bibitem[Dal~Fiume et~al.\ 1998]{1583}
Dal~Fiume, D., et~al. 1998,
\newblock A\&A, 329, L41

\bibitem[Frontera et~al.\ 1997]{1386}
Frontera, F., Costa, E., Dal~Fiume, D., Feroci, M., Nicastro, L., Orlandini,
  M., Palazzi, E., \& Zavattini, G. 1997,
\newblock A\&AS, 122, 357

\bibitem[Frontera \& Dal~Fiume 1989]{1286}
Frontera, F., \& Dal~Fiume, D. 1989,
\newblock in {Proc. 23$^{rd}$ ESLAB Symposium on Two Topics in X--ray
  Astronomy. 1. X--ray Binaries}. (Noordwijk: ESA Publications Division,
  SP--296), 57

\bibitem[Grove et~al. 1995]{375}
Grove, J.E., et~al. 1995,
\newblock ApJ, 438, L25

\bibitem[Guainazzi \& Matteuzzi 1997]{1620}
Guainazzi, M., \& Matteuzzi, A. 1997,
\newblock Technical Report TR--11, BeppoSAX Scientific Data Center

\bibitem[Harding et~al.\ 1983]{304}
Harding, A.K., M{\'e}sz{\'a}ros, P., Kirk, J.K., \& Galloway, D.J. 1983,
\newblock ApJ, 278, 369

\bibitem[Herold 1979]{1219}
Herold, H. 1979,
\newblock Phys.Rev., D19, 2868

\bibitem[Kii et~al.\ 1986]{658}
Kii, T., Hayakawa, S., Nagase, F., Ikegami, T., \& Kawai, N. 1986,
\newblock PASJ, 38, 751

\bibitem[Lammers 1997]{1650}
Lammers, U. 1997,
\newblock {The SAX/LECS Data Analysis System User Manual},
\newblock SAX/LEDA/0010

\bibitem[Makishima \& Mihara 1992]{407}
Makishima, K., \& Mihara, T. 1992,
\newblock in Frontiers of X--ray Astronomy, ed. Y. Tanaka \& K. Koyama
  (Tokyo: Universal Academy Press), 23

\bibitem[Makishima et~al.\ 1990]{61}
Makishima, K., et~al. 1990,
\newblock ApJ, 365, L59

\bibitem[Manzo et~al.\ 1997]{1533}
Manzo, G., Giarrusso, S., Santangelo, A., Ciralli, F., Fazio, G., Piraino, S.,
  \& Segreto, A. 1997,
\newblock A\&AS, 122, 341

\bibitem[M{\'e}sz{\'a}ros 1992]{1614}
M{\'e}sz{\'a}ros, P. 1992,
\newblock {High-Energy Radiation from Magnetized Neutron Stars},
\newblock (Chicago: Chicago Univ. Press)

\bibitem[Mihara 1995]{1547}
Mihara, T. 1995,
\newblock Ph.D. thesis, Univ.of Tokyo

\bibitem[Nagel 1981]{306}
Nagel, W. 1981,
\newblock ApJ, 251, 278

\bibitem[Orlandini et~al.\ 1998]{1581}
Orlandini, M., et~al. 1998,
\newblock A\&A, 332, 121

\bibitem[Owens, Oosterbroek, \& Parmar 1997]{1612}
Owens, A., Oosterbroek, T., \& Parmar, A.N. 1997,
\newblock A\&A, 324, L9

\bibitem[Parmar et~al.\ 1997]{1531}
Parmar, A.N., et~al. 1997,
\newblock A\&AS, 122, 309

\bibitem[Pravdo et~al.\ 1979]{617}
Pravdo, S.H., et~al. 1979,
\newblock ApJ, 231, 912

\bibitem[Shinoda et~al.\ 1990]{29}
Shinoda, K., Kii, T., Mitsuda, K., Nagase, F., Tanaka, Y., Makishima, K., \&
  Shibazaki, N. 1990,
\newblock PASJ, 42, L27

\bibitem[Soong et~al.\ 1990]{1172}
Soong, Y., Gruber, D.E., Peterson, L.E., \& Rothschild, R.E. 1990,
\newblock ApJ, 348, 641

\bibitem[Tanaka 1986]{1584}
Tanaka, Y. 1986,
\newblock in {Radiation Hydrodynamics in Stars and Compact Objects}, ed.
  D. Mihalas \& K.H. Winkler (Berlin: Springer), 198

\bibitem[Tr{\"u}mper et~al.\ 1978]{576}
Tr{\"u}mper, J., Pietsch, W., Reppin, C., Voges, W., Staubert, R., \&
  Kendziorra, E. 1978,
\newblock ApJ, 219, L105

\bibitem[Vaughan \& Kitamoto 1997]{1590}
Vaughan, B.A., \& Kitamoto, S. 1997,
\newblock ApJ,
\newblock submitted (astro-ph/9707105)

\bibitem[White, Swank, \& Holt 1983]{303}
White, N.E., Swank, J.H., \& Holt, S.S. 1983,
\newblock ApJ, 270, 711

\end{thebibliography}
\end{document}